\documentclass[journal=jacsat,manuscript=article]{achemso}
\usepackage[version=3]{mhchem} 
\usepackage[T1]{fontenc}
\usepackage{xcolor}
\usepackage[T1]{fontenc} 
\usepackage{graphicx}
\usepackage{hyperref}
\hypersetup{colorlinks,linkcolor={blue},citecolor={blue},urlcolor={blue}}
 
\author{Hillol Kumar Barman}
\affiliation{Department of Physics, Indian Institute of Technology Bombay, Powai, Mumbai 400076, India}
\email{hillol@iitb.ac.in}
\author{Pathik Das}
\email{ms21090@iisermohali.ac.in}
\affiliation{Department of Physical Sciences, Indian Institute of Science Education and Research (IISER) Mohali, Sector 81, SAS Nagar, Mohali, Punjab 140306, India
}
\author{Syed Yunus Ali}
\affiliation{Department of Physics, Indian Institute of Technology Bombay, Powai, Mumbai 400076, India}
\email{syed_24@iitb.ac.in}

\title{Fluctuations in first passage times and utility of resetting protocol in biochemical systems with two-state toggling}


\date{\today}
\begin{document}

\begin{abstract}
Interesting theoretical problems of target search or threshold crossing, formally known as {\it first passage}, often arise in both diffusive transport problems as well as problems of chemical reaction kinetics. We study three systems following different chemical kinetics, and are special as they 
{\it toggle between two states}: (i) a population dynamics of cells with auto-catalytic birth and intermittent toxic chemical-induced forced death, (ii) a bond cluster model representing membrane adhesion to extracellular matrix under a fluctuating load, and (iii) a model of gene transcription with a regulated promoter switching between active and inactive states.  Each of these systems has a target state to attain, which defines a first passage problem -- namely, population becoming extinct, complete membrane detachment, or mRNA count crossing a threshold. We study the fluctuations in first passage time and show that it is interestingly {\it non-monotonic} in all these cases, with increasing strength of bias towards the target. We also study suitable {\it stochastic resetting} protocols to expedite first passage for these systems, and show that there is a re-entrant transition of the efficacy of this protocol in all the three cases, as a function of the bias. The exact analytical condition for these transitions predicted in earlier literature is verified here through simulations.  
\end{abstract}

\section{{Introduction}\label{sec1}}
Stochastic processes abound in nature, and appear in physical, chemical, and biological systems \cite{Gardiner1985HOSM}. In biological systems, stochastic processes may be broadly classified into: (i) {\it spatial transport processes} which include e.g. bacterial run-and-tumble\cite{Berg1972nat}, dynamic instability of microtubules\cite{Dogterom1993prl}, diffusive molecular motor \cite{Julicher1997ROMP,Kolomeisky2007ARPC}, and (ii) {\it chemical reaction kinetics} which include e.g. gene expression, receptor-ligand binding, and population dynamics \cite{Kondev2012PBOC,Schwarz2013RMP,Allen2003AISPWAB}. In this paper we will be interested in various stochastic chemical reaction kinetics. 

Within stochastic processes, the study of first passage problems in which a target state is reached for the first time, has a long history particularly for various variants of random walks  \cite{Chandra1943RMP,Redner2001AGTFP,Metzler2014FPPATA,Majumdar2013AIP}. In biological systems first passage problems arise at various scales -- from intracellular to organismic and population level \cite{Orsanga_2014}. Some examples from recent literature include kinetochore capture by microtubules\cite{Nayak2020prs}, encounter of remote parts of DNA \cite{Zhang2016ARB}, first binding to  sites on DNA \cite{parmar2016nar, Iwahara2021BiChem}, backtrack recovery in transcriptional proofreading \cite{Roldan2016pre}, and protein threshold-level crossing leading to cell lysis \cite{kannoly2020iscir,Rijal2020pre,Rijal2022prl}, pore formation on endosomes \cite{santra2024jcs}, or cell division \cite{Iyer2014prl}. Interesting non-monotonic behaviour of fluctuations of first passage time\cite{Animesh2020SM}, as well as of its mean \cite{Moumita2024SM}, in different disordered environments have been experimentally demonstrated \cite{Friedman2020jpcl}. In this paper, we will study first passage events in systems with chemical reaction kinetics which specifically, toggles between two dynamical states.

Such systems with two-state toggling arise in directed and diffusive transport processes like run-and-tumble \cite{Malakar_2018,Dhar_2019PRE,Evans_2018,Santra_2020}, Brownian motion in flashing potentials \cite{Santra_2021}, motor transport switching between processive motion (on microtubules) and diffusive sojourns (in 3d) \cite{Benichou2011RMP,Nandi2025PRE}. {{Transport with toggling between two distinct diffusive states has been studied \cite{Cao_2001_PRE}.} with Two-state toggling may also arise in systems with reaction kinetics, some examples of which have been given above, and will be studied elaborately in this paper. Certain general characteristics of target search, first passage, and efficacy of stochastic resetting strategy were studied in \cite{Hillol2025pre} -- although examples studied therein were all related to spatial transport -- such that the mathematical description was based on coupled Fokker-Planck equations. In this paper, we demonstrate the applicability of the broad conclusions in \cite{Hillol2025pre} in systems with reaction kinetics, which all follow coupled Master equations for stochastically evolving discrete numbers.   

Stochastic resetting, in which the current state of the system is occasionally set back to its initial state, has been shown to be a useful strategy to minimize mean first passage times (FPT). The strategy was first introduced in \cite{Evans2011prl} for an ordinary diffusion in one dimension. It was shown that long trajectories which otherwise makes the walker stay away from the target, are curtailed by frequent resets at an optimal rate, and mean FPT is minimized. Important fluctuation properties associated with this strategy was studied in \cite{Shlomi2016prl} and the subsequent developments which followed over a decade may be found in \cite{Kundu2024preface}. It has been theoretically studied in the context of animal foraging\cite{APalforaging}, RNA polymerase backtracking and recovery \cite{Roldan2016pre}, cytonemes searching multiple cell targets \cite{Bressloff2020_PRE}, active particle motion \cite{Baouche2025arxiv}, and also experimentally with optical traps \cite{Friedman2020jpcl,Majumdar_2020PRR}, and programable robots \cite{Nitin2024PRXL}. Within Michaelis–Menten reaction scheme \cite{Cao_2017_JPCLett}, occasional unbinding and rebinding of the enzyme from the enzyme–substrate complex is a resetting mechanism as discussed in \cite{Reuveni2014pnas,Reuveni_MM_PRE}. By regulating the unbinding and the reaction restart rate, the average reaction turnover time (a first passage time) may be optimized. Resetting has also been found to be useful in precipitation accumulation \cite{Ali2022jphysa}, in escape over potential barriers \cite{Singh2025chaos}, and other scenarios like periodic potentials and discrete lattice walks \cite{Ghosh2023jcp,Debraj2022jphysa}.

It has been shown that resetting strategy is helpful when fluctuation in FPT without resetting is  high, and not when it is low \cite{Pal2017prl,Ray2019jphysa}. This  criterion becomes useful when bias towards the target compete with resetting to offset its advantage.   In such cases, in systems without state toggling, the general mathematical condition where the advantage of resetting vanishes through a continuous transition was derived through a Landau-like expansion for the mean FPT \cite{Saeed2019pre,Pal2019prr} -- it is given by $CV^2=1$,  where $CV^2 = \rm{Variance/(mean)^2}$ for FPT in the absence of resetting. This condition has proved useful to demarcate the boundary of regions in parameter space across which resetting may or may not be helpful as a strategy \cite{Saeed2019pre,Saeed2020pre,Saeed2022pre,Saeed2023jphysa}. 

For systems with toggling between two states (say $\sigma=$``$+$'' or $\sigma=$``$-$''), the condition $CV^2=1$ is replaced by two new conditions. We note firstly that there are now two mean first passage times  $\langle T_r\rangle^+$ and $\langle T_r\rangle^-$ for initial states being `$+$'' and `$-$'', which are minimized by tuning Poisson resetting rate $r$ at optimal points $r_{\ast}^+$ and $r_{\ast}^-$ respectively. The dynamical transitions at which the efficacy of resetting vanish, are marked by  $r_{\ast}^+\rightarrow 0$ and $r_{\ast}^-\rightarrow 0$, leading to the following conditions of transition:
\begin{eqnarray}
    \big( \langle T \rangle ^{++}\big)^2+ \langle T \rangle ^{-+}(\langle T \rangle ^{+-} + \langle T \rangle ^{++}+ \langle T \rangle ^{--})-\frac{1}{2}( \langle T^2 \rangle ^{++}+ \langle T^2 \rangle ^{-+})=0,\label{e1} 
\end{eqnarray}
corresponding to $\langle T_r\rangle^+$, and 
\begin{eqnarray}
    \big( \langle T \rangle ^{--}\big)^2+ \langle T \rangle ^{+-}(\langle T \rangle ^{-+}+ \langle T \rangle ^{++} + \langle T \rangle ^{--})-\frac{1}{2}( \langle T^2 \rangle ^{+-}+ \langle T^2 \rangle ^{--})=0,\label{e2}
\end{eqnarray}
corresponding to $\langle T_r\rangle^-$. The detailed derivation of these conditions  starting from renewal theory for two-state systems and developing Landau-like expansion for $\langle T_r\rangle^+$ and $\langle T_r\rangle^-$ in $r$, may be found in \cite{Hillol2025pre}. The moments $\langle T \rangle ^{\sigma_f\sigma_i}$ and $\langle T^2 \rangle ^{\sigma_f\sigma_i}$ in Eq.~\ref{e1} and Eq.~\ref{e2} are related to the survival probability $Q_0^{\sigma_f\sigma_i}(t)$  up to time $t$ in the absence of resetting, with the joint condition that the state is $\sigma_i$ at $t=0$ and the state is $\sigma_f$ at $t$ (see Sec.~\nameref{sim_sec}).

In this paper, we study three chemical systems: (A) stochastic population dynamics with autocatalytic birth, with occasional lethal dosage application to cause death, (B) stochastic adhesion and detachment of a  membrane to extracellular matrix, with intermittent loading, and (C) stochastic transcription of mRNA regulated by a promoter which switches between transcriptionally active and inactive states. The details of the models involving two-state toggling, and various results related to them, appear in Sec.~\nameref{model}. The important goals of this investigation are: (i) to study the nature of fluctuations of FPT, and (ii) explore the efficacy of stochastic resetting strategies to minimize mean FPT in the systems. The similarity of our results below among these distinct chemical systems as well as  the systems of diffusive transport studied in \cite{Hillol2025pre}, demonstrate their generality across systems with two-state toggling. We note that analytically exact solutions for the first passage problems in the two-state systems treated in this paper do not exist in the literature and are challenging, as  non-constant transition rates (or effectively non-linear potentials) and coupled Master equations are involved. This is unlike some analytically solvable cases with linear potentials as treated in \cite{Hillol2025pre}. Hence the whole study in this paper is based on simulations -- the numerical details of the calculations may be found in Sec.~\nameref{sim_sec}. We provide concluding remarks in Sec.~\nameref{conclusion}. 

\section{Methods}\label{sim_sec}
\subsection{Simulation Details}
We studied the models in the Sec.~\nameref{model} using Gillespie simulations\cite{gillespie1977jpc}.
We simulated the time evolution of the population size (e.g. mRNA count, bond number or cell number), starting from a chosen initial value. Gillespie simulations are event driven simulations in which stochastic updates of the system are implemented following rates which appear in the Master equations defining the model (e.g. Eq.~(\ref{bdme1}),(\ref{bdme2}),(\ref{bcme1}),(\ref{bcme2}),(\ref{p_plus}) and (\ref{p_minus}) in this paper). Events occur with probability proportional to their rates, and at random times following a Poisson process \cite{gillespie1977jpc}. For our simulations in this work, the different rates are of population increment and decrement, rates of state change between $``+"$ and $``-"$, and resetting rate of the population size.    
Resetting events were implemented by returning the population to its initial size in the birth-death model and the bond-cluster model, and to half of the desired threshold value ($X/2$) in the gene transcription model. First-passage events at which the simulations stop, were recorded when the population reached the target size zero in the birth-death and bond-cluster model, and size $X$ in the gene transcription model. For reliable statistics, all results were averaged over a large ensemble of independent simulation histories (see the numbers below).
The simulations had two different aspects which we call as method 1 and method 2 below.

\subsection{Method 1: Calculation of mean FPT as a function of stochastic resetting rate $r$.}

Gillespie simulations were done using the rates of events which appear in the Master equations corresponding to the different models in the Sec.~\nameref{model}. Simulations were done in the absence of resetting to obtain data in Figures \ref{bd_res}(a), \ref{bc_res}(a), and \ref{f4}(a), as well as Fig.~S1 (Supporting Information). In the presence of non-zero resetting rate $r$, mean FPT was calculated as a function of $r$. 
{The mean FPTs are obtained by recording the total time taken to reach the first-passage event and averaging this quantity over all simulated histories}.
Following the location of the minimum point of $\langle T_r\rangle^-$ at $r_\ast^-$ for various $k_{off}$ and bias strengths (namely $k_d$ or $k_+$ or $k_m$), the parameter space diagram and the inset boxes in Figures \ref{bd_res}(b), \ref{bc_res}(b)  and \ref{f4}(b), were obtained. The points where $r_\ast^-$ becomes zero are shown in empty black circles in these figures-- the precision of these locations were within an error of $\sim 0.02$ which are indicated by error bars. 

For obtaining accurate results we took the intervals $dr$ of resetting rate $r$  to be sufficiently small, and the number of histories ($H$) to obtain the statistics of FPTs to be  sufficiently high. For the birth-death model $dr=0.005$, $H=5 \times 10^7$; the bond-cluster model  $dr=0.001$, $H=10\times 10^7$;  and the mRNA transcription model $dr=0.005$, $H=5 \times 10^7$. 

\subsection{Method 2: Evaluating the analytical conditions for the vanishing transition of optimal resetting rates.}

Eqs.~\ref{e1} and Eq.~\ref{e2} provide the analytical conditions where the optimal resetting rates $r_\ast^+$ and $r_\ast^-$ vanish respectively \cite{Hillol2025pre}. These conditions involve moments $\langle T\rangle^{\sigma_f\sigma_i}$ and $\langle T^2\rangle^{\sigma_f\sigma_i}$ which are related to the joint survival probability $Q_0^{\sigma_f\sigma_i}(t)$ as follows:
\begin{align}
    \langle T\rangle^{\sigma_f\sigma_i}&=\int_0^\infty Q_0^{\sigma_f\sigma_i}(t) dt,\label{int1}\\
    \langle T^2\rangle^{\sigma_f\sigma_i}&=\int_0^\infty 2t Q_0^{\sigma_f\sigma_i}(t) dt.\label{int2}
\end{align}
Here $\sigma_i=\pm$ and $\sigma_f=\pm$ are the states of the system initially and at time $t$, respectively. The quantity $Q_0^{\sigma_f\sigma_i}$ is the probability of survival in absence of resetting up to time $t$, such that the state is $\sigma_i$ at $t=0$ and $\sigma_f$ at $t$. 

Using the Gillespie simulation method 1 above, we first evaluate the four joint probabilities $Q_0^{++}(t)$, $Q_0^{-+}(t)$, $Q_0^{+-}(t)$, and $Q_0^{--}(t)$ as a function of $t$, for every model discussed in Sec.~\nameref{model}. Typically, we used $\sim 10^7$ histories for every parameter combination to obtain these joint probabilities. We used a fixed time interval $dt=10^{-4}$ for obtaining the time dependence of the joint survival probabilities. Since the time progression in the underlying Gillespie simulation is stochastic, we take the state $\sigma_f(t)$ to be the one at the last Gillespie update before time $t$. Integrating the joint survival probabilities numerically using the formulas in Eq.~\ref{int1} and \ref{int2}, we evaluated the desired moments necessary for Eqs.~\ref{e1} and \ref{e2}. The parameter values of $k_{off}$ and bias strengths at which Eq.~\ref{e2} was satisfied, were used to plot the solid black lines in Figures \ref{bd_res}(b), \ref{bc_res}(b), and \ref{f4}(b).

\section{Models and Results}\label{model}

In this section, we introduce the models of three chemical systems in the following three subsections, along with the respective Master equations which describe their dynamics. The results of the calculations on the statistics of FPT in the presence and absence of resetting, are provided in the respective subsection corresponding to each model.   

\subsection{(A) Population dynamics with auto-catalytic birth, forced death, and extinction }
\begin{figure*}
    \centering
    \includegraphics[width=0.95\textwidth]{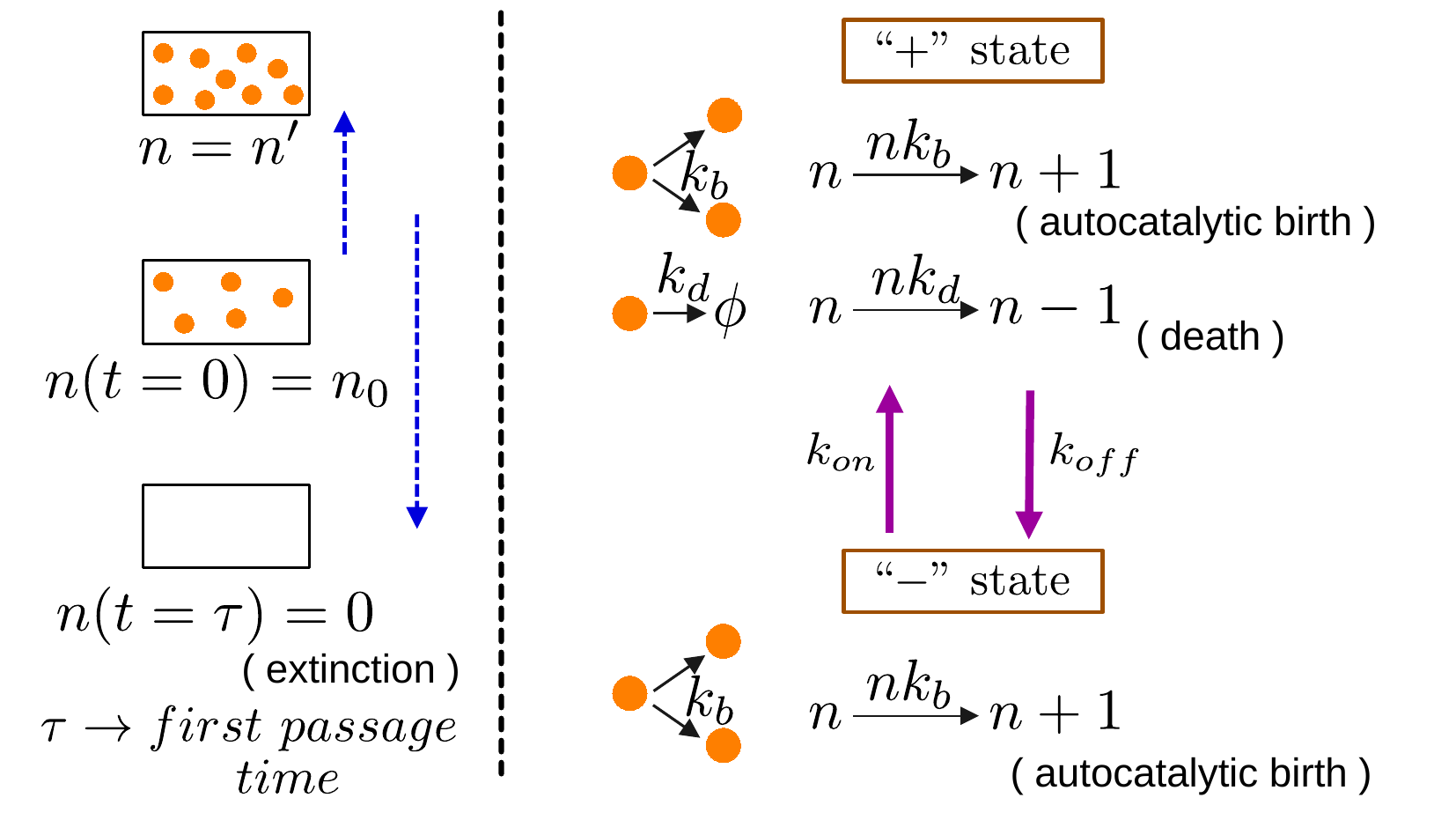}
    \caption{A model of population dynamics. (Left panel) Schematic of the first passage problem showing the initial state (at $t=0$), a state at intermediate time, and the final state  (at $t=\tau$). (Right panel)
    The kinetics in the two states $``+"$ and  $``-"$ with the autocatalytic birth, death, and state toggling are shown with their corresponding rates.}
    \label{bd_sc}
\end{figure*}

Birth-and-death processes play a crucial role in stochastic population dynamics. Autocatalytic birth has been studied in the context of bacterial growth, which has an exponentially growing mean and exponentially large fluctuations \cite{Delbruck1940jcp}; the impact of such fluctuations on pathogen ecologies has been studied \cite{Das_2012}. Auto-catalytic growth has also been used to model growth of protein biomass in the context of cell division \cite{Iyer2014spnas}, in  protocell models of metabolism-first theories of origin of life \cite{Bissette_2015,Xavier2020,Ameta2023CC}, and cancer cell growth \cite{Boman2020}.  Death process on the other hand, is similar to radioactive decay and well studied in the literature \cite{Gardiner1985HOSM}. Generalized birth-death processes with time-dependent rates and spatial diffusion have been studied \cite{Kendall_1948,Adke_1964}. First passage problems in birth-death processes have also been of interest, and in particular, for the event of {\it extinction} where the population size goes to zero for the first time \cite{Adke_1964,MASUDA_1988,Allen2003AISPWAB,Orsanga_2014}. Measures of extinction times in population biology, provide insights into species persistence, ecosystem stability, and drug induced extermination of pathogen or diseased cells. 

In our model (shown in the right side of Fig. \ref{bd_sc}), we consider bacteria (or any other cells) growing auto-catalytically ($A \rightarrow A + A$). The growth rate is $n k_b$ to have a birth ($ n \rightarrow n+1$), where $n$ is the instantaneous population size. The system can be in two states. In the $``-"$ state, there is only autocatalytic birth, which always takes $n$ away from $0$, i.e. extinction.  We ignore a separate basal death rate which may be present in reality. In the $``+"$ state, in addition to autocatalytic birth, there is `forced death' by infusing lethal chemical agents like antibiotics or anticancer drugs such that $n \rightarrow n -1$ at rate $nk_d$ --- note that this process pushes $n$ towards extinction ($n=0$). The switch from the $``-" \rightarrow ``+"$ state happens at rate $k_{on}$ while the reverse transition happens at rate $k_{off}$.   
The domain of the population size is $n\in [0,\infty)$, as there is no upper bound due to any finite system size. The probabilities $P_+(n,t)$ and $P_-(n,t)$ of  the population size $n$, in  $``+"$ and $``-"$ states respectively, satisfy the Master equations: 

\begin{align}
\frac{\partial P_+(n,t)}{\partial t} &= (n-1)k_bP_+(n-1,t)+ (n+1)k_dP_+(n+1,t) \nonumber \\
&\qquad \qquad \qquad \quad\ \ -n(k_b+k_d)P_+(n,t)    +k_{on}P_-(n,t) -k_{off} P_+(n,t),\label{bdme1}\\
\frac{\partial P_-(n,t)}{\partial t} &= (n-1)k_bP_-(n-1,t) -nk_bP_-(n,t)+k_{off}P_+(n,t)-k_{on}P_-(n,t).\label{bdme2}
\end{align}

\begin{figure*}
    \centering
    \includegraphics[width=1\textwidth]{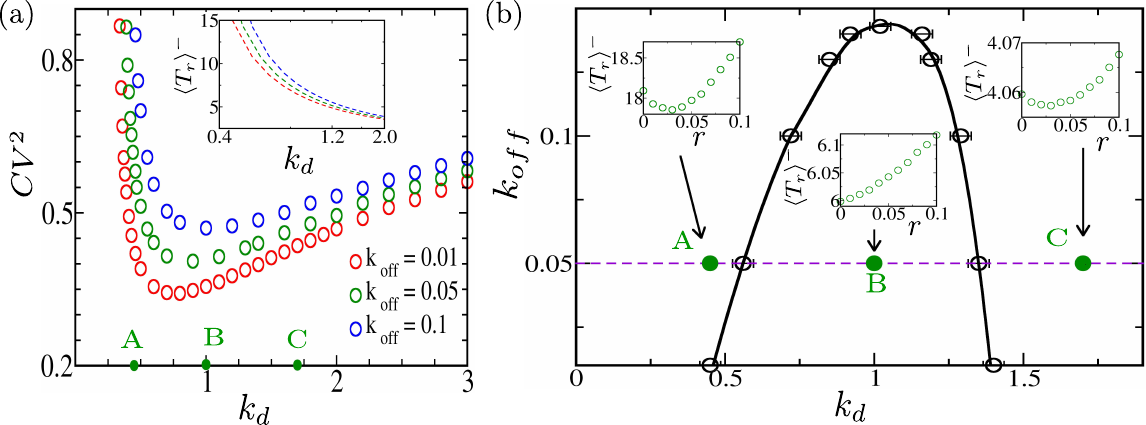}
    \caption{(a) $CV^2$ as a function of $k_d$ shows an U-shape for three different $k_{off}$ indicated with labels and colors. Here $k_{on}=0.51$, $n_0=5$, and $k_b=0.3$.
    Inset: mean FPT $\langle T_r\rangle^-$ decreases monotonically with $k_d$ (for the three different $k_{off}$ according to the colors). The points A, B, and C marked on the $k_+$ axis (for $k_{off}=0.05$) correspond to high, low, and high $CV^2$, respectively.
   (b) The solid black line obtained using Eq.~\ref{e2} demarcate the boundary between the regions $r_{\ast}^- \ne 0$ and  $r_{\ast}^- = 0$.  The mean FPT $\langle T_r\rangle^-$ vs $r$ at the three points A, B, and C (at $k_{off}=0.05$) are shown in separate boxes. The empty black circles with error bars represent the points where  $r_{\ast}^-$ vanishes, obtained through Gillespie simulations.}
    \label{bd_res}
\end{figure*}

In our model, starting from a population size $n_0$, the size $n$ evolves stochastically following the kinetics defined above, and first passage happens when the population goes extinct at time $\tau$ (see the left panel of Fig.~\ref{bd_sc}).
In  Fig.~\ref{bd_res} we present the statistics related to the FPT of extinction ($\tau$) for the model. In the inset of Fig.~\ref{bd_res}(a), we show that the mean FPT $\langle T_r\rangle^-$ (starting from the growing state $``-"$) monotonically decreases as the death rate $k_d$ in the chemical infused state is increased. This is the usual expectation, even if there was a single $``+"$ state without toggling where birth and death is present.  On the other hand, we see in Fig.~\ref{bd_res}(a) that the fluctuations of  FPT, captured through squared coefficient of variation $CV^2$,  shows a non-monotonic behaviour. At small $k_d$ higher fluctuations are expected, as time spent up to extinction is typically large, and with finite birth rate $k_b$, noise enhances.  At large $k_d$, the role of toggling between states gains importance. Although in most of the cases the system goes extinct quickly due to large $k_d$ (see small $\langle T_r\rangle^-$), a fraction of configurations get delayed due to residence in and visits to the ``$-$'' state, contributing to high $CV^2$ -- see the magnitudes get higher with increasing $k_{off}$ (Fig.~\ref{bd_res}(a)). In Fig.~S1 (Supporting Information), the plots of $CV^2$ vs $k_d$ for the starting state to be $``+"$, 
shows that the U-shape is absent without toggling (i.e. $k_{off} = 0$), and thus  supports the claim that at high $k_d$ fluctuations get higher due to state toggling (i.e. $k_{off} > 0$).  We will show below in this model (as well as in the other two models in the next subsections), that the efficacy of resetting strategy overlaps roughly with the regions of high FPT fluctuations ($CV^2$).  

Can extinction be expedited (i.e. mean FPT is minimized) by an experimental strategy? We define a stochastic resetting protocol in the system whereby, at random times (at rate $r$), some cells are introduced or killed to return the population size to its initial value $n=n_0$. The Master equations Eq.~\ref{bdme1} and Eq.~\ref{bdme2} are augmented with the terms $+r\delta_{n,n_0}-rP_+(n,t)$ and $+r\delta_{n,n_0}-rP_-(n,t)$ respectively.
In  Fig.~\ref{bd_res}(b), as $k_d$ is increased, we see that the mean FPT $\langle T_r\rangle^-$ may be optimized with a finite resetting rate $r=r_*^-$ at points A and C (where $CV^2$ of FPT is high). In contrast, $r_*^-=0$ at point B (where $CV^2$ of FPT is low), implying that resetting the population size at this point would not help in making the extinction process quicker. Yet, $CV^2 > 1$ is no longer the condition for finding efficient resetting, as in systems without toggling.  We instead follow the exact analytical Eq.~\ref{e2} to plot the boundaries (see Sec.~\nameref{sim_sec}) between the resetting-efficient and resetting-redundant regions in the  $k_d-k_{off}$ parameter space (see the solid line in Fig.~\ref{bd_res}(b)). Direct Gillespie simulations (see Sec.~\nameref{sim_sec}) give the resetting advantage vanishing boundary (see the empty black circles in Fig.~\ref{bd_res}(b)), which coincident with the analytically predicted solid line.  These transitions show that the resetting is effective in aiding extinction only at low and high death rates $k_d$, and not at its intermediate values.

\subsection{(B) Bond-cluster model for membrane adhesion and detachment, under intermittent force}

\begin{figure*}[!ht]
    \centering
    \includegraphics[width=0.9\textwidth]{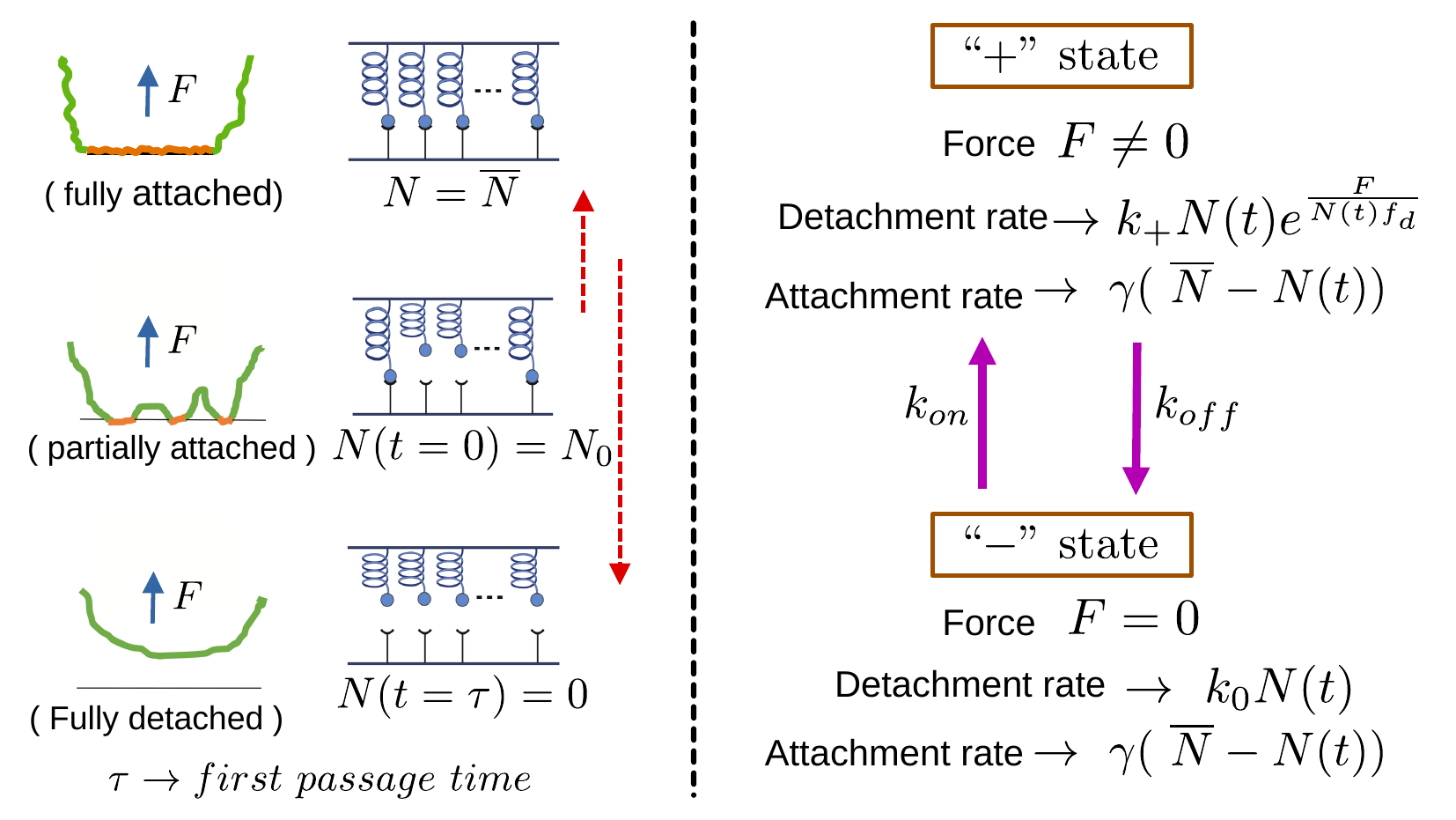}
    \caption{ Membrane adhesion and bond cluster model. (Left panel) Schematic of the first passage problem showing the initial partially attached membrane (at $t=0$), the limiting case of fully attached membrane, and the final state of fully detached membrane at $t=\tau$.
    (Right panel) The kinetics in the two states $``+"$ (with nonzero force) and  $``-"$ (with zero force) are shown with the corresponding attachment, detachment, and state toggling rates.}
    \label{bc_sc}
\end{figure*}
Adhesion interactions through bond formation of cell surface receptors like integrin with ligands (e.g. fibronectin)  on the extracellular matrix (ECM), play a critical role in the mechanical stability of focal adhesions, driving cell migration  \cite{Bachman2015AWC, Gumbiner1996cell,Discher2005Sci, Schwarz2013RMP}. Such adhesion interactions have been studied in cell-mimetic systems \cite{Sackmann2014SM}. At a focal adhesion, integrin proteins mediate between the internal cellular actin-myosin network (which often flows) and the ECM, with other proteins like talin playing a crucial role in the dynamic stability \cite{Sackmann2014SM}. The complexity is partially captured through different models.  
Effective membrane models with continuum Helfrich free energy, along with additional energetic terms, have been studied for cell adhesion  \cite{Discher2005Sci,Sunnick2012_PRE,Vink2013_SM}. Discrete lattice gas models \cite{Dharan2014_JCP} have also been studied.   

Motivated by the above biophysical context, in this work, we have focused on certain theoretical bond-cluster models in the literature  \cite{Bell1978Sci, Seifert2000prl,Erdmann2004jcp}, which ignore  spatial organization and study the simple aspect of fluctuating number of attached and detached bonds within the cluster. There is some experimental support for such parallel bonds clusters and their rupture \cite{Noy2005Bioj,Sulchek2005pnas}.   
A schematic model of bond-cluster is shown in the left panel of Fig.~\ref{bc_sc}. Such a  bond-cluster model recently studied membrane  detachment as a {\it first passage} problem, but with fixed force $F$ \cite{Dasana2020prr}. Yet the detachment of the bonds from the ECM is influenced often by dynamic loading force $F$-- in fact time varying cyclic forces at focal adhesion have been reported in experiments  \cite{Plotnikov2012cell,stricker2011bioj,Plotnikov2013COCB}. These two sets of literature motivate us to study the first passage problem of membrane detachment in the bond-cluster model (Fig.~\ref{bc_sc}) in which additionally the force stochastically varies leading to a 2-state toggling scenario.

The bond cluster model instantaneously has $N(t)$ attached bonds (such that $0 \leq N \leq \overline{N}$), in which the attachment rate is $\gamma (\overline{N} - N(t))$. The loading force can fluctuate stochastically between two values: when $F\ne 0$ (in the  $``+"$ state), the detachment rate is $ k_+ N(t) \exp\left(\frac{F}{N(t) f_d}\right) $, while for $F=0$ in the  $``-"$ state, the detachment rate is $k_0 N(t)$ (see right panel of Fig.~\ref{bc_sc}). The constant $f_d$ has dimension of force. The system toggles from the force-free state to the forced state ($``-" \rightarrow{ ``+"}$) at rate $k_{on}$, and from $``+" \rightarrow{ ``-"}$ state at rate $k_{off}$. The Master equation for this system with probability $P_+(N,t)$ and $P_-(N,t)$ of $N$ attached bonds in the $``+"$ and $``-"$ state respectively are given by:
\begin{align}
\frac{\partial P_+(N,t)}{\partial t} &= \alpha_1^{(N-1)}P_+(N-1,t)+\alpha_2^{(N+1)}P_+(N+1,t) -(\alpha_1^{(N)}+\alpha_2^{(N)})P_+(N,t)\nonumber\\&\qquad \qquad \qquad \qquad  +k_{on}P_-(N,t) -k_{off} P_+(N,t),\label{bcme1}\\
\frac{\partial P_-(N,t)}{\partial t} &=\alpha_1^{(N-1)}P_-(N-1,t)+k_0(N+1)P_-(N+1,t) -(\alpha_1^{(N)}+k_0N)P_-(N,t)\nonumber\\& \qquad \qquad \qquad \qquad  +k_{off} P_+(N,t)-k_{on}P_-(N,t),\label{bcme2}
\end{align}
where $\alpha_1^{(N)}=\gamma (\overline{N}-N)$ and $\alpha_2^{(N)}=k_+N(t) \exp\left(\frac{F}{N(t) f_d}\right)$.

We define the first passage event as the complete detachment of the membrane from the ECM \cite{Dasana2020prr}. As shown in Fig.~\ref{bc_sc}, starting from a partially attached membrane ($N=N_0$), the number of attached bonds fluctuates and eventually goes to $0$ -- the time $\tau$ of this event defines the FPT (see the left panel of Fig.~\ref{bc_sc}). In this problem, we are interested to vary the strength of the detachment rate constant $k_+$ in the  $``+"$ state and study how the statistics of FPT are affected by it. In the inset of Fig.~\ref{bc_res}(a), we show that the mean FPT decreases monotonically with  $k_+$  -- as expected, average time is shorter with more efficient detachment. In contrast, the  $CV^2$ of FPT (Fig.~\ref{bc_res}(a)) goes from a high to a low and then to a high value. 
Large fluctuations at small $k_+$ is expected, as attachments compete with weakly forced detachments to lead to many noisy excursions away from the target ($N=0$). In contrast, the impact of state toggling manifests at large $k_+$ -- detachment happens super efficiently via the $``+"$ state in many cases, while in some, prolonged delays happen due to residence in the  $``-"$ state, giving rise to the diversity in FPT. This is clarified in Fig.~\ref{bc_res}(a) where $CV^2$ rises with $k_{off}$ at high $k_+$. Moreover, in Fig.~S1 (Supporting Information) for the $CV^2$ corresponding to FPT starting from the  $``+"$ state, we see no U-shape at $k_{off}=0$, confirming the active influence of state toggling at high $k_+$ to enhance the $CV^2$.  

\begin{figure*}
     \centering   
     \includegraphics[width=1\textwidth]{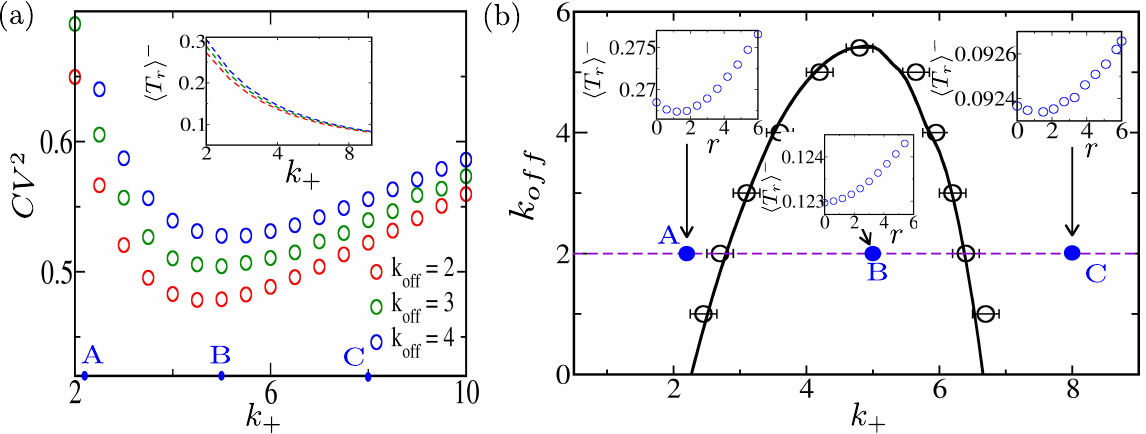}
\caption{(a) $CV^2$ as a function of $k_+$ shows an U-shape for three different $k_{off}$ indicated with labels and colors. Here $k_{on}=20$, $F=3.5$, $N_0=3$, $\overline{N}=10$, $k_0=0.2$, $\gamma=1$,and $f_d=1$.
Inset: mean FPT $\langle T_r\rangle^-$ decreases monotonically with $k_+$ (for the three different $k_{off}$ according to the colors). The points A, B, and C marked on the $k_+$ axis (for $k_{off}=4$) correspond to high, low, and high $CV^2$, respectively.
(b) The solid black line obtained using Eq.~\ref{e2} demarcate the boundary between the regions $r_{\ast}^- \ne 0$ and  $r_{\ast}^- = 0$.  The mean FPT $\langle T_r\rangle^-$ vs $r$ at the three points A, B, and C (at $k_{off}=4$) are shown in separate boxes. The empty black circles with error bars represent the points where  $r_{\ast}^-$ vanishes, obtained through Gillespie simulations.
}
\label{bc_res}
 \end{figure*}

With the aim to minimize the mean FPT, we introduce the following mechanical intervention strategy in which at random times selected at rate $r$, some bonds are either attached or detached externally to bring back the number $N$ to the initial value $N_0$. The Master equations acquires additional terms for resetting:  $+r\delta_{N,N_0}-rP_+(N,t)$ and $+r\delta_{N,N_0}-rP_-(N,t)$ in Eq.~\ref{bcme1} and Eq.~\ref{bcme2} respectively.  In Fig.~\ref{bc_res}(b), we see that the expectation to minimize FPT is borne out in regions where FPT fluctuations are high (e.g., points A and C), but not where it is low (e.g., at point B). Plotting $\langle T_r\rangle^-$ versus $r$, we see that at point $A$ and $C$ there are finite optimal values $r_*^- \neq 0$, while at point B the $r_*^- = 0$ (Fig.~\ref{bc_res}(b)). Thus, we have a first transition from a regime where the resetting protocol gives an advantage, to another where it is not useful, and then again return to a regime where it is useful through a second transition. The exact boundary of these two transitions in the $k_{off} - k_+$ plane plotted from Eq.~\ref{e2} (shown in solid line in Fig.~\ref{bc_res}(b)), matches with the points obtained through direct Gillespie simulations (see \nameref{sim_sec}). The calculation thus reveals that for a certain intermediate range of detachment strength ($k_+$), a mechanical strategy to occasionally reset attached bond numbers does not lower the mean FPT of membrane detachment from the ECM. The strategy would although help at small or large values of $k_+$.

\subsection{(C) Gene transcription for a regulated promoter, and mRNA threshold crossing}
\begin{figure*}[!ht]
    \centering    \includegraphics[width=0.95\textwidth]{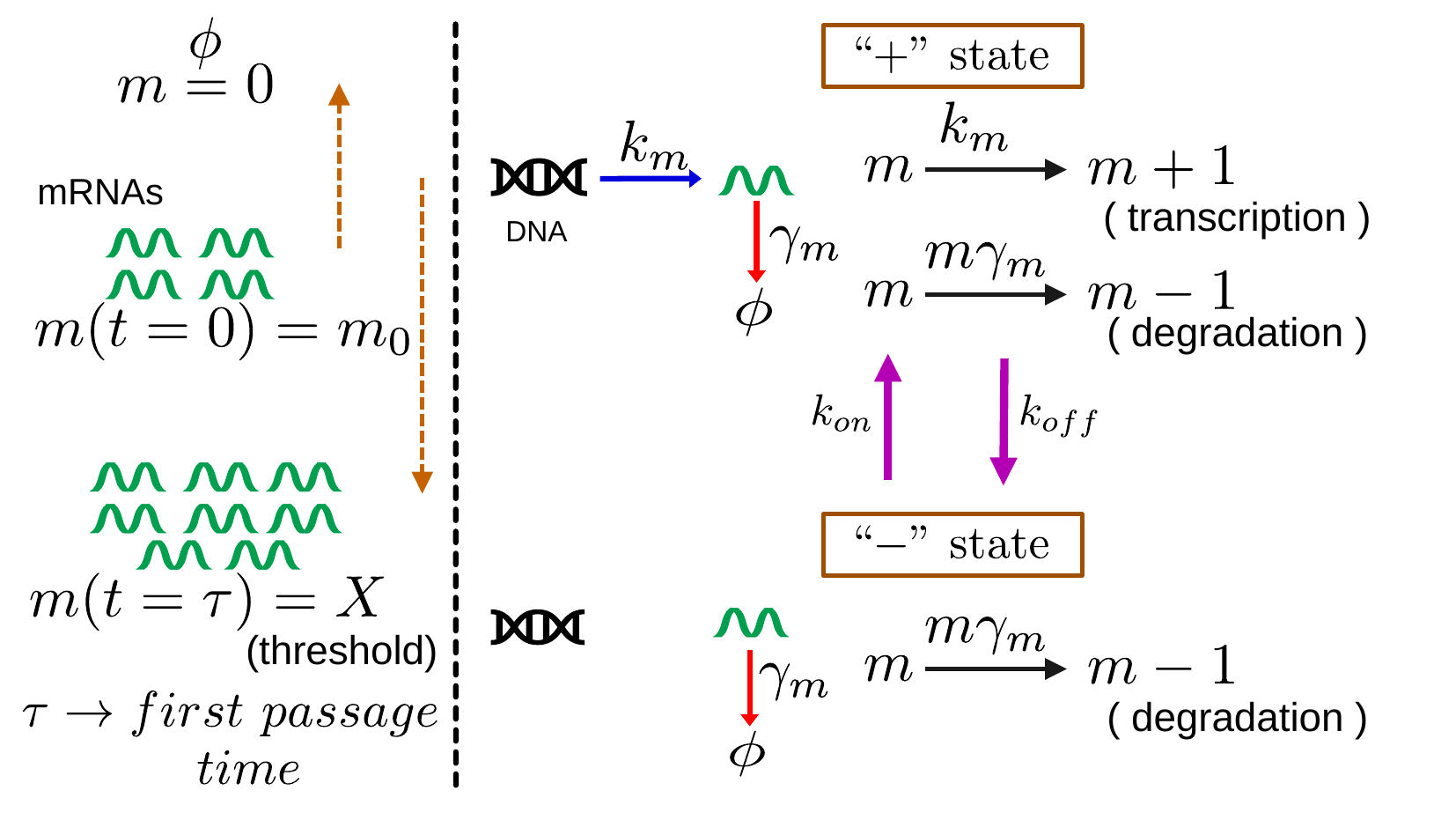}
    \caption{ Model of mRNA transcription.
     (Left panel) Schematic of the first passage problem showing the initial mRNA count (at $t=0$), the limiting case of zero transcript, and the final state where the mRNA count has reached the threshold $X$ (at $t=\tau$).
    (Right panel) The kinetics in the transcriptionally active  $``+"$ state,
    inactive  $``-"$ state, and toggling between two with suitable rates is depicted.   
    }
    \label{f3}
\end{figure*}
Gene expression is inherently stochastic \cite{thattai2001pnas, Elowitz2002sci,choubey2013meth} -- there is large cell-to-cell variability of copy numbers of mRNA and protein. The variability arises due to multiple causes, including transcriptional and post-transcriptional regulation, leading to irregular and bursty production of mRNAs and proteins \cite{Raser2004sci,Raj2006Bursting,singer2014MOlcell,shahrezaei2008pnas,wang2023jcp}, as well as noise in cell cycle and cell division \cite{Huh2011natgen,Soltani2016ploscmp,Ali2025bioarxiv}. Here we would study the standard two-state regulated promoter model, in which there is a toggling between transcriptionally active (``$+$'') and inactive (``$-$'') states \cite{Raj2008cell,suter2011sci,sanchez2011pcb,singh2013ieee}. The switch from active to inactive state may be due to the occlusion of regions of the promoter by the nucleosomes \cite{parmar2016nar},  engineered ligand-inducible transcription factors \cite{Ott2018Natcommun}, or by altering the shape of the promoter \cite{O’Halloran2015Sci}.

The model of mRNA transcription that we study is shown on the right side of Fig.~\ref{f3}. In the active state (``$+$''), gene transcription rate is $k_m$ leading to mRNA count $m\rightarrow{m+1}$,  while in the inactive state (``$-$''), there is no production of mRNA. In both states, mRNA degrades ($m\rightarrow{m-1}$) with a rate $m\gamma_m$. The state ${``+"}\rightarrow{``-"}$ with rate $k_{\mathrm{off}}$ and the switch $``-"\rightarrow{``+"}$ happens with rate $k_{\mathrm{on}}$. This kinetics leads to a stochastic evolution of the mRNA count in a cell, which may be mathematically described by the following Master equation:

\begin{align}
\frac{\partial P_+(m,t)}{\partial t} &=k_m P_+(m-1,t) + \gamma_m (m+1) P_+(m+1,t)
-(k_m+m\gamma_m)P_+(m,t)\nonumber\\ 
& \quad +k_{\mathrm{on}} P_-(m,t) - k_{\mathrm{off}} P_+(m,t),\label{p_plus}\\
\frac{\partial P_-(m,t)}{\partial t} &=\gamma_m [(m+1) P_-(m+1,t) - m P_-(m,t)] +k_{\mathrm{off}} P_+(m,t) - k_{\mathrm{on}} P_-(m,t) . \label{p_minus}
\end{align}

Here $P_+(m,t)$ and $P_-(m,t)$ denote the probability of having $m$ mRNA molecules at time $t$ in the $``+"$ and $``-"$ states, respectively. 
We would now proceed to define a first passage problem in the model of transcription presented above. There is considerable theoretical interest in first passage times (FPT) for protein threshold-level crossing \cite{raj2017pnas,Rijal2020pre,Rijal2022prl,Iyer2014prl}. Such first passage processes are known to regulate timings of crucial cellular events like cell lysis on infection by virus \cite{singh2014jrsci,kannoly2020iscir}, pore formation in endosomes containing toxin-secreting bacterium \cite{santra2024jcs}, and cell division \cite{suckjoon2019cb, Iyer2014spnas}. Recently, protein threshold crossing in the presence of post-transcriptional regulation has been studied \cite{Biswas2020pre,Biswas2021epje,Ali2025pre}. The first passage problem of threshold crossing of mRNA has been of interest in experiment \cite{mukherji2011natgen}, and has been studied through approximate theories \cite{biswas2019epje}. No analytically exact solutions exist in the literature for FPT in the two-state transcriptional model (Fig.~\ref{f3}), although exact results are known in the constitutive production model when the promoter is always active \cite{Gardiner1985HOSM,Rijal2022prl}. In this work, we study the first passage problem for the regulated transcription model, in which starting from mRNA count $m_0$, the number fluctuates and eventually reaches a threshold $m=X$ at FPT $\tau$ (see left side of  Fig.~\ref{f3}). Thus, the domain of $m \in [0,X]$.

\begin{figure*}
    \centering
    \includegraphics[width=1\textwidth]{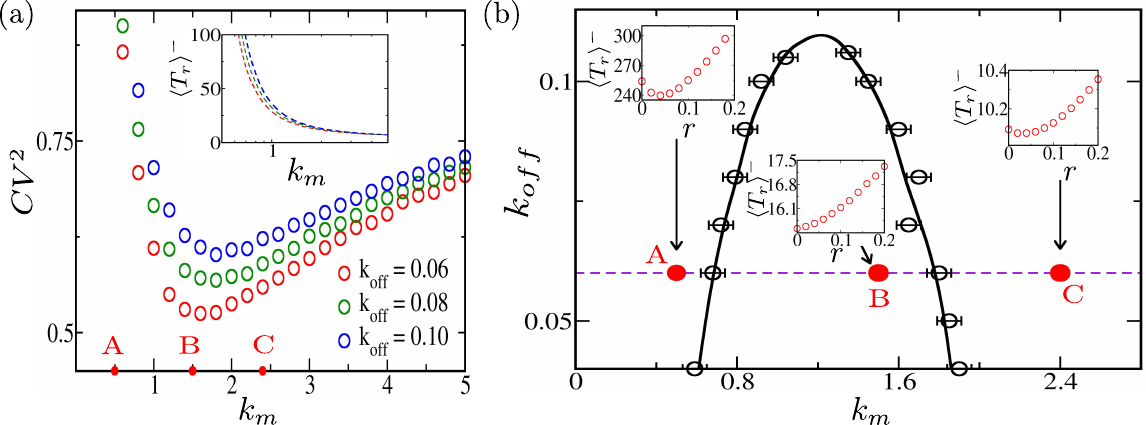}
    \caption{(a) $CV^2$ as a function of $k_m$ shows an U-shape for three different $k_{off}$ indicated with labels and colors. Here $k_{on}=0.2$,  $\gamma_m=0.1$, $X=10$, and $m_0=5$. 
    Inset: mean FPT $\langle T_r\rangle^-$ decreases monotonically with $k_m$ (for the three different $k_{off}$ according to the colors). The points A, B, and C marked on the $k_m$ axis (for $k_{off}=0.06$) correspond to high, low, and high $CV^2$, respectively.
    (b) The solid black line obtained using Eq.~\ref{e2} demarcate the boundary between the regions $r_{\ast}^- \ne 0$ and  $r_{\ast}^- = 0$.  The mean FPT $\langle T_r\rangle^-$ vs $r$ at the three points A, B, and C (at $k_{off}=0.06$) are shown in separate boxes. The empty black circles with error bars represent the points where  $r_{\ast}^-$ vanishes, obtained through Gillespie simulations.}   
    \label{f4}
\end{figure*}

In  Fig.~\ref{f4} we present the results related to the FPT problem of mRNA threshold crossing defined above. In the inset of Fig.~\ref{f4}(a), we show the mean FPT  $\langle T_r\rangle^-$ (for the initial state being inactive $``-"$) monotonically decreases as the transcription rate $k_m$ is increased. Thus, in spite of state toggling, just like constitutive production, the threshold is attained faster with a higher production rate of mRNA. In contrast to this behaviour, the fluctuations of FPT characterized by $CV^2$ have a non-monotonic shape in  Fig.~\ref{f4}(a). Higher fluctuations are expected at small $k_m$ as passage times are long and degradation of mRNA during prolonged residence makes threshold crossing more noisy -- this role of finite degradation has been discussed in systems without state toggling too \cite{raj2017pnas,Ali2025pre}. The high fluctuations at large $k_m$, on the other hand, are influenced by the presence of two states. Although in most cases when the system escapes to the $``+"$ state, there is very quick passage for large $k_m$ (see low $\langle T_r\rangle^-$), as the system starts from the $``-"$ state and may revisit it with finite likelihood, there may be some instances where $\tau$ get large. Such disparate instances give rise to high $CV^2$. 
Note in Fig.~\ref{f4}(a), $CV^2$ rises at high $k_m$ with higher $k_{off}$ (which is associated with higher residence time in the $``-"$ state). We also show in Fig.~S1 (Supporting Information) that the $CV^2$ with the initial state $``+"$, has no $U$-shape in the absence of state toggling, and the feature emerges and intensifies with finite and higher values of $k_{off}$.   

Although experimentally challenging, we theoretically propose a resetting protocol for the problem in the following way. At random times chosen at a fixed rate $r$, some mRNA molecules are added or removed from the system so that the number $m$ is reset to the initial value $m_0$. Analytically this adds terms $+r\delta_{m,X/2} -rP_+(m,t)$  to Eq.~\ref{p_plus} and $+r\delta_{m,X/2}-rP_-(m,t)$ to Eq.~\ref{p_minus} respectively. 
In  Fig.~\ref{f4}(b) we see that at low $k_m$ (point A), $\langle T_r\rangle^-$ has a minimum at $r= r_{\ast}^-$. At intermediate $k_m$ (e.g at point B) a similar plot shows that $r_{\ast}^-=0$. At high $k_m$ (e.g., point C), a minimum is seen, and $r_{\ast}^-$ is again finite. Thus, the regions of resetting efficacy roughly overlap with regions of high FPT fluctuations (having points A and C). As $r_{\ast}^- \ne 0$ implies a useful resetting strategy to optimize the mean FPT, the above behavior of $r_{\ast}^-$ with $k_m$ indicates a re-entrant transition of the efficacy of the resetting protocol. In Fig.~\ref{f4}(b), we plot the phase boundary in the $k_m-k_{off}$ plane following the analytical Eq.~\ref{e2} with solid line, separating the regions of  $r_{\ast}^- \ne 0$ from  $r_{\ast}^-=0$. In empty black circles, we show the points where  $r_{\ast}^-$ vanishes from direct Gillespie simulation data (see Sec.\nameref{sim_sec}) of $ \langle T_r\rangle^-$ vs $r$ --- they coincide with the solid line within error bars. In summary, we have shown that if mRNA number is reset occasionally, then mRNA threshold crossing times may be minimized by the protocol at low or high transcription rates ($k_m$), but interestingly not at intermediate values.

\section{Conclusions}\label{conclusion}

We have presented a study on ``target search'' (or first passage) in chemical systems which toggle between two dynamical states. As discussed in the Sec.~\nameref{sec1}, first passage problems arise in various contexts in cell biology-- e.g. first capture of kinetochore by moving microtubules, first binding of protein to the promoter of DNA, size threshold controlled cell division, or cell lysis. In chemical kinetics, reaction completion times are regarded as first passage times. The fluctuations of FPT are measured using different quantifiers in the literature namely, Coefficient of variance (CV) \cite{choubey2013meth,Reuveni2014pnas,Reuveni_MM_PRE,Shlomi2016prl,Pal2017prl,Ray2019jphysa,Rijal2020pre,kannoly2020iscir,Rijal2022prl,Hillol2025pre,Ali2025pre,Ali2025bioarxiv}, Fano factor\cite{thattai2001pnas,Animesh2020SM}, or Poisson indicator\cite{Cao_2017_JPCB} -- interestingly, often these studies have found mon-monotonic behaviour of these fluctuation indicators as a function of relevant control parameters in the respective problems.

We also investigated the role of the stochastic resetting strategy in this paper.
In the Sec.~\nameref{sec1} we have discussed an extended literature on this topic, and its utility in reaction schemes like Michaelis–Menten, where reaction turnover times may be optimized by suitable unbinding rates of enzyme-substrate complex. In this paper, we have studied reset strategies by theoretically suggesting the removal and addition of molecules or bonds, which help to reach a target population size more efficiently. The exact details of the protocols and their consequences have been discussed in the last paragraphs of each model in the Sec.~\nameref{model}. We found that in these systems with two-state toggling, the success of such strategies non-trivially depends on the parameter of bias towards the target. 

While the theoretical study of first passage and utility of resetting strategy in both spatial transport processes and chemical reactions is an old subject, what is interesting are some characteristic features recently reported \cite{Hillol2025pre} and highlighted in this study for systems with two-state toggling.  
To bring out the generality, we considered three quite diverse examples -- the first involved a well known population dynamics model interrupted by dosage infusion to aim at population extinction, the second involved intermittent loading of a membrane adhered to ECM to aim for its detachment, and lastly the well known regulated active-inactive transcription model aiming at threshold-level crossing of the mRNA transcript. Interestingly, these examples showed some common features which align with the findings in the transport processes reported earlier \cite{Hillol2025pre}. The two-state systems seem to have non-trivial variation of fluctuations of first passage time, and associated limitations on efficacy of resetting strategy, as a function of the bias towards the target. 

We found the following results which seem to be generic: (i) While by tuning up the bias  (e.g. death rate, detachment rate constant, transcription rate) towards the target, the average FPT can be lowered, the fluctuations of FPT do not have a simple monotonic behaviour. Instead, fluctuations can be minimized for an intermediate value of the bias strength. Thus if {\it the aim is to optimize not just the mean FPT but the fluctuations too}, then there seems to be {\it a preferred choice of bias strength}. This reminds of similar results in problems of transport with two-state toggling:  run-and-tumble particle having a preferred speed, or motors detaching and reattaching to microtubules having a preferred processive speed  \cite{Hillol2025pre}. (ii) If a reset protocol is introduced which occasionally takes back the number of the chemical species to its initial value, then the mean time of passage may be lowered, at low or high values of bias strength, but not at intermediate ones. The regions of efficient resetting roughly overlap with the regions of high FPT fluctuations, but there is no exact criterion involving the $CV^2$.  (iii) What exactly demarcates the regimes of successful resetting strategy from those where it is harmful is not the condition of $CV^2 = 1$ (as in systems without state toggling), but the new conditions in Eq.~\ref{e1} and \ref{e2} \cite{Hillol2025pre}. In this paper, we checked the validity of these conditions determining the phase boundaries by comparing with direct Gillespie simulations.  The main achievement was thus to show that these results are general and common to chemical systems, just like the diffusive transport processes studied earlier.  

Analytical exact calculations would be good to have in future for these problems with coupled Master equations and non-constant (and sometimes non-linear) variable dependent transitions rates. This paper will hopefully inspire experimental works -- even if stochastic resetting protocols we have proposed are hard to practically achieve, the non-monotonic fluctuation property which imply preferred bias strengths, may be possibly studied. The bias strengths can be varied by tuning the toxicity of chemicals in the birth-death problem, by choosing the chemical composition of ECM in the membrane detachment case, and by choosing biological variants with different transcription rates for the problem of regulated transcription.   We thus hope that this work would motivate further studies in chemical systems with two-state toggling.

In this paper, we have focused on models having discrete jump processes of numbers (e.g., population size, attached bond number, mRNA copy number), governed by the Master equations. But biophysical scenarios may sometimes demand continuous descriptions of stochastically evolving smooth processes. Models for such systems, with similar questions as we investigate in this paper, have been studied in the past, and further scope is there to study them in future.    
In living cells, cell volume dynamically changes due to cell growth and division. 
In biophysical scenarios where cell growth is important, often mRNA or protein  concentration based treatments rather than discrete number, were preferred \cite{Xie2006_PRL,Rijal2022prl}. In the future, models with two-state promoter toggling and with continuous mRNA concentrations may be studied for threshold crossing problems as possible extensions of this work. Similarly, limitations of the discrete bond cluster model that we studied, like its neglect of spatial organization and hydrodynamic flows, may be addressed in the future by starting with some aspects of the continuum models which already exist in the literature  \cite{Sackmann2014SM,Sunnick2012_PRE,Vink2013_SM}. Time dependent first passage questions need to be posed within such continuum models. We hope the insights we obtained from this work may serve as some guide for those future works.

\begin{acknowledgement}
H.K.B. acknowledges the fellowship support from University Grants Commission (UGC), India, and support of IIT Bombay. S.Y.A. acknowledges the institute post-doctoral fellowship provided by IIT Bombay. The authors sincerely thank Profs. Dibyendu Das and Amitabha Nandi for engaging discussions, as well as for reading and suggesting edits to the manuscript. P.D. thanks the Soft Matter group at IIT Bombay for internship opportunity, and guidance throughout the work.

\end{acknowledgement}

\begin{suppinfo}
Figures related to fluctuations of FPT, starting initially from the ``$+$'' is shown.
\end{suppinfo}

\section{Conflict of interest}
The authors have no conflicts to disclose.

\bibliography{bibfile}

\end{document}


\section{Non-monotonicity of fluctuation of FPT, starting from the ``$+$'' state }
\begin{figure*}
    \centering
    \includegraphics[width=1\textwidth]{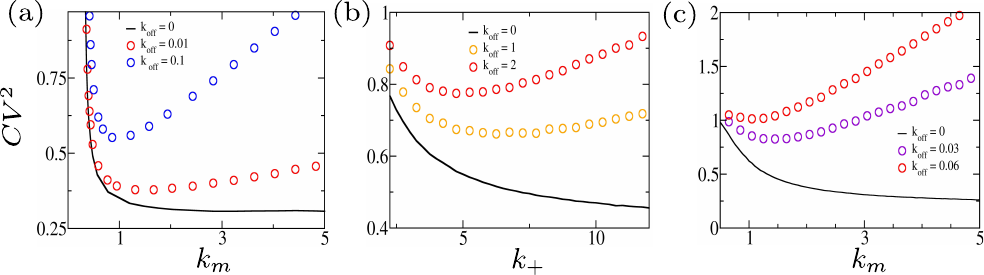}
    \caption{$CV^2$ of FPT for the initial state being ``$+$'', is plotted against the bias strength of the three models studied in the main manuscript. In all the graphs, the case of $k_{off}=0$ (no toggling), is shown in solid black line -- we see that there is no non-monotonic behaviour of $CV^2$. With state toggling and the rise of $k_{off}$, the fluctuations rise and the U-shape is seen. (a) In the population dynamics model, the bias is the death rate per cell $k_d$. Here $k_b=0.3$, $n_0=5$, and $k_{on}=0.51$.(b) For the membrane-ECM adhesion model, the bias strength is the detachment rate constant $k_+$. Here $k_{on}=20$, $F=3.5$, $N_0=3$, $\bar{N}=10$, $k_0=0.2$, $\gamma=1$,and $f_d=1$. (c) In the regulated gene transcription model, the bias is the transcription rate $k_m$. The parameters used are $\gamma=0.1$, $k_{on}=0.2$, and $X=10$.}
    \label{S1}
\end{figure*}